\documentclass[fdp,a4paper,fleqn]{w-art}

\usepackage{times,cite,w-thm}
 \numberwithin{equation}{section}
\theoremstyle{plain}

\theoremstyle{definition}

\usepackage[english]{babel}
\usepackage[]{graphicx}

\usepackage{psfrag}
\usepackage{wasysym}

\newcommand{\be}{\begin{equation}}
\newcommand{\ee}{\end{equation}}

 \newcommand{\Z}{\mathbf{Z}}
 \newcommand{\PP}{\mathbf{P}}

 \newcommand{\half}{\frac 1 2}

 \newcommand{\ket}[1]{|#1\rangle}

\begin{document}
\DOIsuffix{theDOIsuffix}
\Volume{55}
\Month{01}
\Year{2007}
\pagespan{1}{}



\title[Quantum 2-Body Hamiltonian for TCC]{Quantum 2-Body Hamiltonian for Topological Color Codes}


\author[F. Author]{H. Bombin\inst{1}}%
\address[\inst{1}]{Department of Physics, 
Massachusetts Institute of Technology, 
Cambridge, Massachusetts 02139, USA}
\author[S. Author]{M. Kargarian\inst{2}}
\address[\inst{2}]{Physics Department, Sharif University of Technology, Tehran 11155-9161, Iran}
\author[T. Author]{M.A. Martin-Delgado\inst{3}%
\footnote{Invited Lecturer at the Scala Conference 2009.}
}
\address[\inst{3}]{Departamento de F\'{i}sica Te\'orica I,
Universidad Complutense, 28040 Madrid, Spain}
   \dedicatory{Dedicated to Prof. Paolo Tombesi 
on occasion of his seventieth birthday.}
\begin{abstract}
We introduce a two-body quantum Hamiltonian model with spins-$\half$ 
located on the vertices of a 2D spatial lattice. The model exhibits an
exact topological degeneracy  in all coupling regimes. This is a 
remarkable non-perturbative effect.
The model has a  $\Z_2\times \Z_2$ gauge group symmetry and 
string-net integrals of motion.
There exists a gapped phase in which the low-energy sector reproduces 
an effective topological color code model. 
High energy excitations fall into three families of anyonic fermions 
that turn out to be  strongly interacting. All these, and more,
are new features not present in honeycomb lattice models like Kitaev model.
\end{abstract}
\maketitle                   




 \renewcommand{\leftmark}
 {H.Bombin, M.Kargarian, M.A. Martin-Delgado: Quantum 2-Body Hamiltonian for TCC}

 \tableofcontents  

\section{Introduction}
\label{sect_I}

The content of our work \cite{2-bodyTCC,2-bodyLarge} fits naturally into the
topics covered during the Scala Conference 2009.
In fact, the acronym Scala
refers to ``Scalable Quantum Computing with Light and Atoms''
and among its many objectives, we may select the following
two major goals:

\noindent i/ to achieve scalable quantum computation;

\noindent ii/ to perform quantum simulations with light and atoms.

We present a new quantum 2-body Hamiltonian on a 2D lattice 
with results that follows the twofold motivation concerning those topics.
This is so because, on one hand, the Hamiltonian system that we
introduce is 
able to reproduce the quantum computational properties of the
topological color codes (TCC) \cite{topologicalClifford,topo3D,tetraUQC} 
at a non-pertubative level. This is an important step towards obtaining
topological protection against decoherence in the quest for scalability.
On the other hand, the fact that the interactions in the Hamiltonian 
appear as 2-body spin (or qubit) terms makes it more suitable for 
its realization by means of a quantum simulation based on light and atoms.

One of the several reasons  for being interested in the
experimental implementation of this
Hamiltonian system is because it exhibits exotic quantum phases of matter
known as topological orders, some of its distinctive features being the 
existence of anyons.

In our everyday 3D world, we only deal with fermions and bosons.
Thus, exchanging twice a pair of particles is a topologically trivial operation.
In 2D this is no longer true, and particles with other statistics are possible: 
anyons. When the difference is just a phase, the anyons are called abelian.
Anyons are a signature of topological order (TO) \cite{wen_book,wen_fqh}, 
and there are others as well:
\begin{itemize}
\item Topological degeneracy of the ground state subspace (GS).

\item  Gapped excited states: localized quasiparticles, anyons.

\item Edge states.

\end{itemize}
\noindent etc.

But where do we find topological orders? These quantum phases of matter
are difficult to find.
If we are lucky, we may find them 
on existing physical systems such as the quantum Hall effect.
But we can also engineer suitable quantum Hamiltonian models, e.g., 
using polar molecules on optical lattices \cite{zoller05,jiang08,ol_review},
or by some other means.

\section{Topological Stabilizers: Toric Codes and Color Codes}
\label{sect_II}

Some of the simplest quantum Hamiltonian models with topological order 
can be obtained from the formalism local stabilizer codes borrowed
from quantum error correction \cite{gottesman96} in quantum information.
These are spin-$\half$ local models of the form
\be
H = -\sum_i S_i, \ \ 
S_i\in \PP_n:=\langle i, \sigma^x_1, \sigma^z_1,\ldots,
\sigma^x_n, \sigma^z_n  \rangle.
\label{stabilizer_Hamiltonian}
\ee
where the stabilizer operators $S_i$ constitute an abelian subgroup
of the Pauli group $\PP_n$ of $n$ qubits, generated by the Pauli matrices
except $-1$. The ground state is a stabilizer code since it satisfy the
condition
\be
S_i\ket{{\rm GS}}=\ket{{\rm GS}}, \forall i, 
\ee
and the excitated states of $H$ are gapped, and correspond to error syndromes
form the quantum information perspective
\be
S_i\ket{\Psi}=-\ket{\Psi}.
\ee

\begin{figure}[h]
\begin{center}
\includegraphics[width=8 cm]{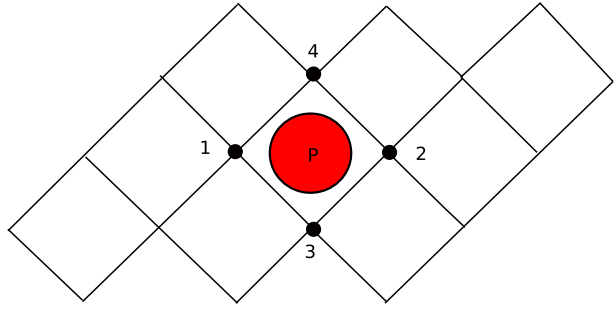}
\caption{\label{toric_code} The square lattice for the
toric code and the plaquette structure of its stabilizers.}
\end{center}
\end{figure}

The seminal example of topological stabilizer codes is the toric code 
\cite{kitaev97}. In a square lattice, we place a spin-$\half$ system 
at each vertex. 
There is one stabilizer $A_p$ per plaquette, see Fig.\ref{toric_code}:
\be
H=-\sum_p A_p, \ \ A_p:= \sigma^x_1 \sigma^x_2 \sigma^z_3 \sigma^z_4
\label{toric_Hamiltonian}
\ee
There exist two kinds of basic excitations.
To label them, we have to color  the plaquettes in black and white,
as in a chessboard. 
There exist a total of three nontrivial topological charges:
two of them are  boson excitations, one per each violation of plaquette;
and the other one is a fermion excitation.  
Let us recall that these boson and fermion charges
are defined in the group $\Z_2$, not the $U(1)$ of electromagnetism.
Excitations come into pairs as the end-points of string configurations.
This is so when the surface is compact without boundaries.

\begin{figure}[h]
\begin{center}
\includegraphics[width=7 cm]{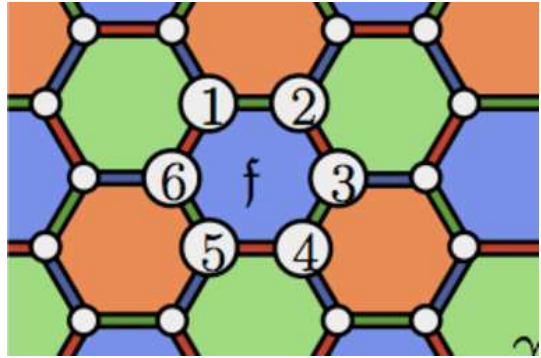}
\caption{\label{color_code} The hexagonal lattice 
is an example of 3-colorable lattice by faces, and also by edges.
A topological color code can be defined on it by associating
two stabilizer operators for each plaquette \eqref{color_stabilizers}.}
\end{center}
\end{figure}

Topological color codes (TCC) are another relevant example
of topological stabilizer codes, with enhanced computational capabilities
\cite{topologicalClifford,topo3D,tetraUQC}. 
In particular, they allow the transversal implementation of Clifford 
quantum operations.
The simplest lattice to construct them is a honeycomb lattice, 
see Fig.~\ref{color_code}, 
where we place a spin-$\half$ system at each vertex.
There are two stabilizer operators per plaquette:
\begin{equation}
\begin{split}
B_p^x&=\sigma^x_1 \sigma^x_2 \sigma^x_3 \sigma^x_4 \sigma^x_5 \sigma^x_6,\\
B_p^y&=\sigma^y_1 \sigma^y_2 \sigma^y_3 \sigma^y_4 \sigma^y_5 \sigma^y_6,
\end{split}
\label{color_stabilizers}
\end{equation}
\be
H_{\rm cc} = -\sum_p (B^x_p + B^y_p).
\label{color_code_Hamiltonian}
\ee

There exist six kinds of basic excitations.
To label them, we first label the plaquettes with three colors:
Notice that the lattice is 3-valent and has 3-colorable plaquettes. 
We call such lattices 2-colexes \cite{topo3D}.
One can define color codes in any 2-colex embedded in an arbitrary
surface.
There exist a total of 15 nontrivial topological charges:
each family of fermions is closed under fusion, and fermions 
from different families have trivial mutual statistics.

\section{Quantum Lattice Hamiltonian with 2-Body Interactions
 for Color Codes}
\label{sect_III}

In nature, we find that interactions are usually 2-body 
interactions. This is because interactions between particles are mediated
by exchange bosons that carry the interactions (electromagnetic, phononic, etc.)
between two particles.

The problem that arises is that for topological models, like the toric codes
and color codes, their Hamiltonians have many-body terms 
\eqref{toric_Hamiltonian},\eqref{color_code_Hamiltonian}. 
This could only
achieved by finding some exotic quantum phase of nature, like FQHE, or
by artificially engineering them somehow.

Here, we shall follow another route: try to find a 2-body Hamiltonian
on a certain 2D lattice such that it exhibits the type of topological
order found in toric codes and color codes. In this way, their physical
implementation looks more accessible.

In fact, Kitaev \cite{honeycomb} 
introduced a 2-body model in the honeycomb lattice 
that gives rise to an effective toric code model in one of its phases.
It is a 2-body spin-$\half$ model in a honeycomb lattice with
one spin per vertex, and simulations based on optical lattices
have been proposed \cite{duan03}.

The model features plaquette and strings constants of motion.
Furthermore, it is exactly solvable, a property that is related
to the  3-valency of the lattice where it is defined.
It shows emerging free fermions in the honeycomb lattice.
If a magnetic field is added, it contains a non-abelian topological 
phase (although not enough for universal quantum computation).

Interestingly enough, 
another regime of the model gives rise to a 4-body model, which is
precisely an effective toric code model.
A natural question arises: Can we get something similar for color codes?
We give a positive answer in what follows.

\subsection{The Model}

\begin{figure}[h]
\begin{center}
\includegraphics[width=10 cm]{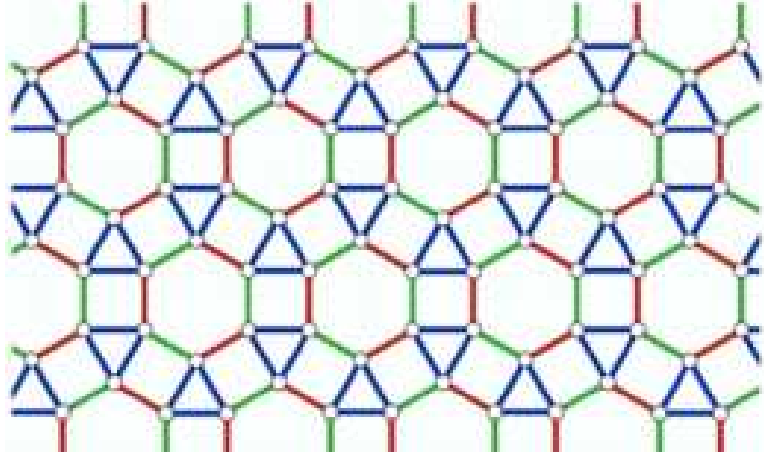}
\caption{\label{ruby_lattice} A lattice with coordination
number 4 where the 2-body quantum lattice Hamiltonian
for the color codes is defined according to spin-spin interactions
coded by the colors of the links, as in \eqref{2-bodyTCC}.}
\end{center}
\end{figure}

It is a 2-body spin-1/2 model in a 'ruby' lattice as shown in 
Fig.\ref{ruby_lattice}. 
We place one spin per vertex.
Links come in 3 colors, each color representing a different interaction.

\be
H = \sum_{\langle i,j\rangle} J_w \sigma^w_i \sigma^w_j, \ \ \ \ \
w = \begin{cases}
x, & \text{ red links} \\
y, & \text{ green links} \\
z, & \text{ blue links} 
\end{cases}
\label{2-bodyTCC}
\ee

For a suitable coupling regime, this model gives rise to an effective color code model. Furthermore, it exhibits new features, 
many of them not present in honeycomb-like models:

\begin{itemize}

\item Exact topological degeneracy in all coupling regimes 
($4^g$ for genus $g$ surfaces).

\item String-net integrals of motion.

\item Emergence of 3 families of strongly interacting fermions 
with semionic mutual statistics. 

\item $\Z_2\times\Z_2$ gauge symmetry. 
Each family of fermions sees a different $\Z_2$ gauge subgroup.

\end{itemize}

\subsection{Integrals of Motion}

\begin{figure}[h]
\begin{center}
\includegraphics[width=10 cm]{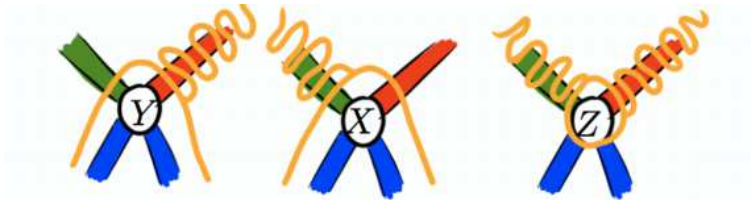}
\caption{\label{plaquette_IOM} A diagrammatic representation
of the local structure of the integrals of motion of the 
2-body Hamiltonian \eqref{color_code_Hamiltonian}. The colored
links represent different spin-spin interactions.}
\end{center}
\end{figure}

We can construct integrals of motion (IOM), $I\in\PP_n, [H_{\rm cc},I] = 0$,
following a pattern of rules assigned to the vertices of the lattice,
as shown in Fig.\ref{plaquette_IOM}. These rules are constructed to 
attach a Pauli operator of type $\sigma^x_i$, $\sigma^y_i$ or $\sigma^z_i$
to each of the vertices $i$. The lines around the vertices, either wavy lines
or direct lines, are pictured in order to join them along paths of vertices
in the lattice that will ultimately  
translate into products of Pauli operators, which will become IOMs. 
Clearly, $\sigma^z_i$ operators are distinguished from the rest.
Therefore, c represent the local structure of the
IOMs of our 2-body Hamiltonian \eqref{color_code_Hamiltonian}.
We will illustrate them with several examples of increasing complexity.

\begin{figure}[h]
\begin{center}
\includegraphics[width=11 cm]{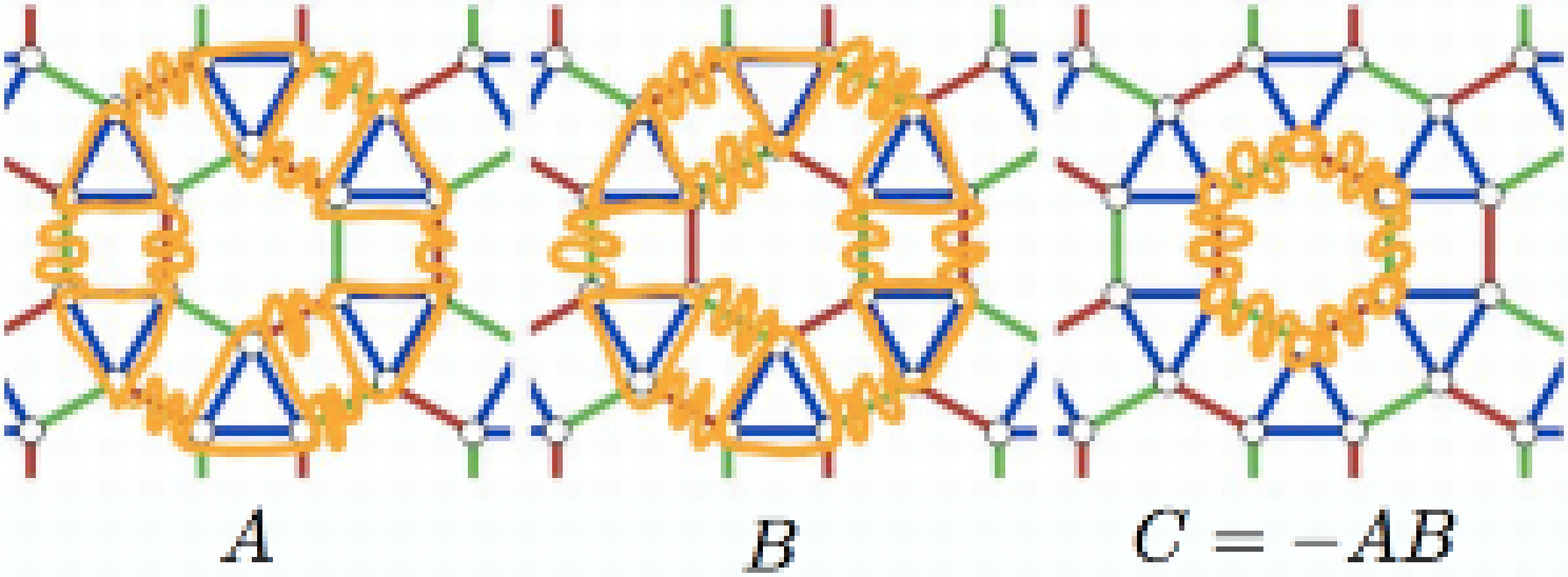}
\caption{\label{plaquette_IOM_Independent} 
Schematic drawing of the plaquette IOMs according to the
local rules in Fig.\ref{plaquette_IOM}. There are 3 IOMs 
denoted as $A,B,C$, but only 2 of them are independent.
This corresponds to the symmetry  $\Z_2\times\Z_2$ of the model.}
\end{center}
\end{figure}

Let us start by constructing the elementary plaquette IOM as 
shown in Fig.\ref{plaquette_IOM_Independent}. They are denoted
as $I=A,B,C$. They are closed since they have not endpoints left.
Using the Pauli algebra, it is immediate to check that they satisfy
$C=-AB$. Thus, there exist 2 independent IOMs per plaquette: 
this is the $\Z_2\times\Z_2$ local symmetry of the model Hamiltonian
\eqref{color_code_Hamiltonian}.

\begin{figure}[h]
\begin{center}
\includegraphics[width=7 cm]{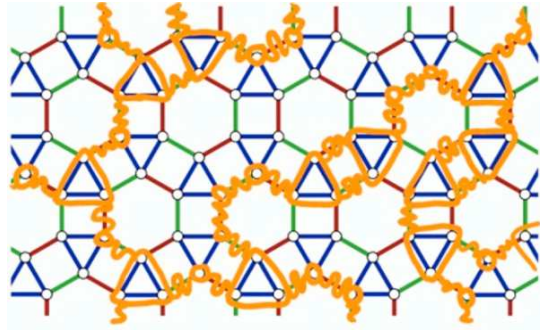}
\caption{\label{stringnet_IOM} 
An example of a stringnet IOM. Notice the presence of branching points
located around blue triangles of the lattice. This is a remarkable difference
with respect to honeycomb models like the Kitaev model.}
\end{center}
\end{figure}

The most general configuration that we may have 
is shown in Fig.\ref{stringnet_IOM}. We call them stringnets IOM
since in the context of our model, they can be thought of as
the stringnets introduced to characterize topological orders \cite{levinwen05}.
The key feature of these IOMs is the presence of branching points
located at the blue triangles of the lattice. This is remarkable
and it is absent in honeycomb 2-body models like the Kitaev model.
When the sringnets IOMs are defined on a simply connected piece of 
lattice they are products of plaquette operators.
More generally, they can be topologically non-trivial and 
independent of plaquette operators.

\begin{figure}[h]
\begin{center}
\includegraphics[width=12 cm]{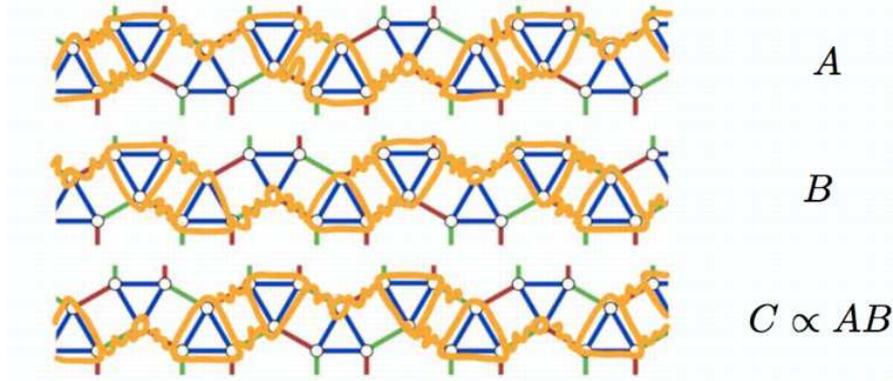}
\caption{\label{strings_IOM} 
Examples of standard string configuration of IOMs, i.e., without
branching points. For each path, we can in principle make 3 different
assignments of IOMs, but again only 2 of them are independent as with
plaquette IOMs.
This is another manifestation of the $\Z_2\times\Z_2$ symmetry  of the model.}
\end{center}
\end{figure}

As a special case of IOM we have string configurations, i.e., 
paths without branching points. Some examples are shown in 
Fig.\ref{strings_IOM}. They may be open or closed, depending on
whether they have endpoints or not, respectively. 
Strings IOM are easier to analyze.
Stringnets IOM are products of strings IOM.
For a given path, there exist 3 different string IOM.
These are denoted as $A,B,C$ in Fig.\ref{strings_IOM}.
Again, using Pauli algebra we get that
only two of them are independent, as with the plaquette IOMs.
To distinguish properly the three types we have to color the lattice.
Strings are then red, green or blue. This is closely related to the
topological color code \cite{2-bodyTCC,2-bodyLarge,topologicalClifford}.

\subsection{Connection with 2-Colexes}

From the previous discussion on IOMs, we have already seen a connection
with the topological color codes. We can make this more quantitative.
In fact, there is a regime of coupling constants in which one of the
phases of the 2-body Hamiltonian reproduces the TCC many-body structure
and physics.

\begin{figure}[h]
\begin{center}
\includegraphics[width=12 cm]{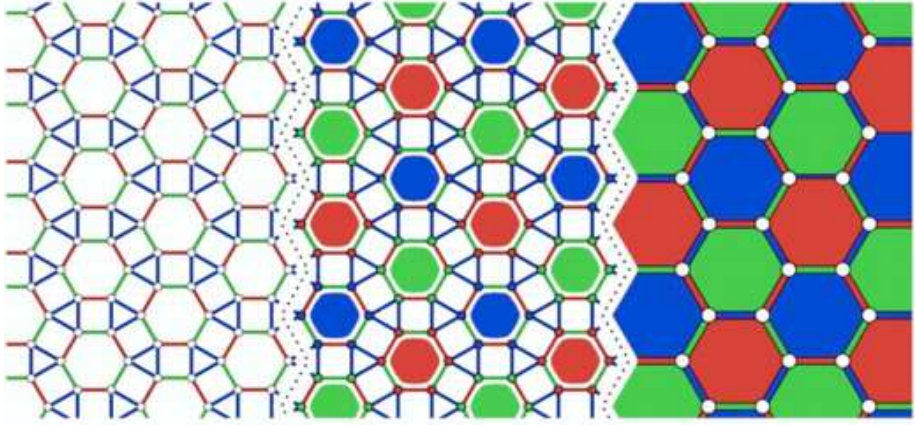}
\caption{\label{transition_to_2Colex} 
The three stages showing the emergence of the topological color code:
(left) the original lattice for the 2-body Hamiltonian 
\eqref{color_code_Hamiltonian}. The color in the links denote
the type of spin-spin interactions; (middle) a different coloring of 
the lattice is introduced based on the property that the hexagons 
are 3-colorable; (right) the hexagonal lattice
obtained by shrinking to a point the blue triangles of the original
lattice, which become sites in the final hexagonal lattice. 
This corresponds to the strong coupling limit in 
\eqref{strong_coupling}.}
\end{center}
\end{figure}

The emergence of the topological color code is beautifully pictured in
Fig.\ref{transition_to_2Colex}. This corresponds to the following set
of couplings in the original 2-body Hamiltonian \eqref{color_code_Hamiltonian}:
\be
J_z = \frac{1}{4}, \ \ J_x, J_y > 0, \ \ J_x, J_y \ll J_z.
\label{strong_coupling}
\ee
This is a strong coupling limit in $J_z$. Geometrically, it corresponds
to shrinking the blue triangles of the original lattice into points,
which will be referred as sites of a new emerging lattice,
see Fig.\ref{transition_to_2Colex} (left). Recall that the blue links of 
these triangles represent spin-spin interactions of $\sigma^z$ type.
Motivated by this strong coupling limit, it is convenient to give another
different coloring to the lattice which will make the transition towards
the hexagonal lattice more transparent. Thus, we realize that 
the hexagons and vertices of the model are 3-colorable, 
see Fig.\ref{transition_to_2Colex} (middle):
if we regard blue triangles as the sites of a new lattice, we get a honeycomb
lattice, see Fig.\ref{transition_to_2Colex} (right).
In fact, the model could be defined for any other 2-colex,
not necessarily an hexagonal lattice.

The topological color code effectively emerges in this coupling regime.
This can be seen using degenerate perturbation theory in the Green
function formalism \cite{2-bodyLarge}. Originally, this method was
applied to the Kitaev model on the honeycomb lattice \cite{honeycomb}.
Alternatively, it is possible to use the PCUTs approach
(Perturbative Continuous Unitary Transformations) \cite{PCUTS}.
This is inspired by the RG method based on unitary transformation
introduced by Wegner (the Similarity RG method) \cite{wegner94}.
Originally, the PCUTs method was applied to the Kitaev model 
\cite{honeycomb_bosonization}. The application of this method 
to our model is based on mapping
from the original spins on the blue triangles to hardcore bosons
with spin. An operator $Q$ serves to count the number of hardcore bosons
by sectors. 
At a given perturbative order, the method 
produces an effective Hamiltonian such that
\be
[H_{\rm eff},Q] = 0.
\ee
We are specially interested in the low-energy $Q=0$ sector, 
where high-energy excitations are not present.
In this sector, only effective spin degrees of freedom are relevant.
Up to a constant, the effective $Q=0$ Hamiltonian at 9th perturbative order 
is \cite{2-bodyTCC}:

\be
H_{\rm eff} = -\sum_p (k_x B^x_p + k_y B^y_p + k_z B^x_p B^y_p)
\label{effective_Hamiltonian}
\ee

\be
k_z = \frac{3}{8} |J_xJ_y|^3 + O(J^7), \ \ 
\frac{k_x}{|J_y|^3} = \frac{k_y}{|J_z|^3} = \frac{55489}{13824} |J_xJ_y|^3.
\ee
This is a color code in the honeycomb lattice of effective spins! 
The ground state is the vortex free sector.
Excitations are vortices. They are gapped and localized at plaquettes.
Higher order terms are products of vortex operators. 
This gives rise to short-range vortex interactions.

At this point, it is illustrative to bring about an analogy with the
fractional quantum Hall effect. In FQHE, the physical degrees of freedom
are electrons under a very strong magnetic field perpendicular to the plane
where the electrons move. The magnetic field is so strong that typically
only the charge of the electrons matter, since their spins are fully polarized.
Originally, the interactions among electrons are 2-body interactions:
the Coulomb electric force. It is possible to write a Hamiltonian based
on these interactions that describe the system from first principles.
However, it is known that a better description is possible:
the Laughlin wave function \cite{laughlin83}. Despite this is an effective
description of the system, it captures the whole new physics of the
electronic system under these extreme circumstances.
Interestingly enough, the Laughlin wave function is not an eigenstate of
the original Hamiltonian based on Coulomb 2-body interactions.
Instead, it is an eigenstate of a Hamiltonian with many-body 
interacting terms. This is the analogy.
The strong coupling limit \eqref{strong_coupling} of our model
corresponds to the extreme regime of the electronic system in the FQHE.
Our original 2-body Hamiltonian \eqref{2-bodyTCC} is like the 
Coulomb Hamiltonian, while the effective topological color code
Hamiltonian \eqref{effective_Hamiltonian} with many-body terms
have eigenstates which play the role of Laughlin wave functions.
In fact, both systems are examples of topological orders
in strongly correlated systems.
However, there is a nice difference: our model retains properties
of the topological color codes at a non-perturbative level because
of the existence of the IOMs, which are exact in all coupling regimes.

\begin{figure}[h]
\begin{center}
\includegraphics[width=9 cm]{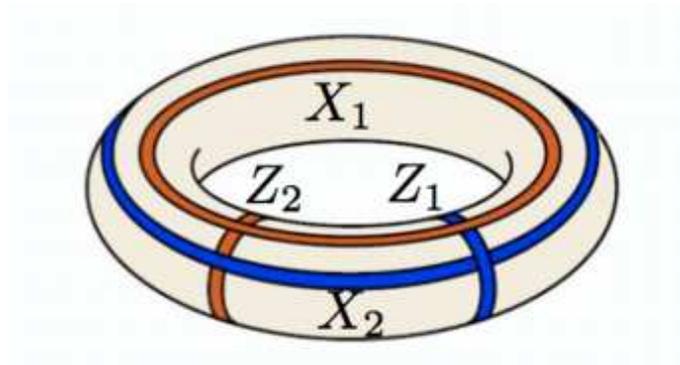}
\caption{\label{color_torus} 
A basis of colored string operators of $X$- and $Z$-type
for a topological color code on a 1-torus.}
\end{center}
\end{figure}

Once we are in the phase corresponding to the topological color code,
then it is possible to relate the original string and stringnets IOMs
with the corresponding configurations in the color code.
Thus, for example, a blue string is composed of blue links and so on
and so forth.
String IOMs of different color that cross once anticommute among each other.
This feature is not available in honeycomb-like models.
If we embed the original 2-body lattice model into a genus 1-torus,
then we shall obtain a TCC in that torus. In that case,
we can choose 4 independent string IOMs that form the algebra of 
Pauli operators on 2 qubits: $\{X_i,Z_i\}, i=1,2$, namely,

\begin{equation}
\begin{split}
[Z_1,Z_2] = 0,\ & [X_1,X_2] = 0, \\
[Z_1,X_2] = 0,\ & [Z_2,X_1] = 0, \\
\end{split}
\end{equation}
\begin{equation}
\begin{split}
\{Z_1,X_1\} = 0,\ & \{Z_2,X_2\} = 0, \\
\end{split}
\end{equation}
\begin{equation}
\begin{split}
X^2_1=1=Z^2_1,\ &  X^2_2=1=Z^2_2. 
\end{split}
\end{equation}
A possible choice for the configurations of these colored string operators
that code the logical qubits is shown in Fig.\ref{color_torus}.

This implies an exact 4-fold degeneracy. 
More generally, in a surface of genus $g$ we find a $4g$-fold 
topological degeneracy.

\section{Conclusions and Future Work}
\label{sect_conclusions}

We have introduced a two-body spin-1/2 model in a ruby lattice
\cite{2-bodyTCC,2-bodyLarge}, see Fig.\ref{ruby_lattice}.
The model exhibits an exact topological degeneracy in all coupling regimes.
Using a bosonic mapping, it is possible to discuss
the emergence of strongly interacting anyonic fermions.
A particular coupling regime gives rise to an effective model
which is a topological color code.

We have shown that the new model exhibit enough novel interesting and
relevant properties so as to justify further research. Some of these
possible lines of study are as follows:

We have only studied a particular phase of the system,
although we are able to study non-perturbative effects as well.
The fact that all phases show a topological degeneracy 
anticipates a rich phase diagram.
In this regard, one may explicitly brake the color symmetry that 
the model exhibits and still keep the features that we have discussed.
It would be particularly interesting to check whether any 
of the phases displays non-abelian anyons. 
The model has many integrals of motion, although not enough to
make it exactly solvable. This becomes another appealing feature of
the model since other methods of study, like numerical simulations
and experimental realizations will help to give a complete 
understanding of all its phases.

\begin{acknowledgement}
We thank the editors of this Special Issue for their kind
invitation to present our written contribution
for the festschrift celebrating Paolo Tombesi's 70th birthday.
MAMD also thanks the Program and the Local organizing committees
for their invitation to lecture at the 
International Scala Conference 2009 on ``Scalable Quantum Computing with Light and Atoms''.
  We acknowledge financial support
 from a PFI grant of EJ-GV, DGS grants under contracts,
FIS2006-04885, and the ESF INSTANS 2005-10.
\end{acknowledgement}


\begin{thebibliography}{99}

\bibitem{2-bodyTCC}
H. Bombin, M. Kargarian, M.A. Martin-Delgado,
``Interacting Anyonic Fermions in a Two-Body `Color Code' Model'';
arXiv:0811.0911.

\bibitem{2-bodyLarge}
M. Kargarian, H. Bombin, M.A. Martin-Delgado;
``Topological Color Codes and Two-Body Quantum Lattice Hamiltonians'';
Submitted to the Special Issue on "Quantum Information and Many-Body Theory", 
New Journal of Physics. Editors: M.B. Plenio, J. Eisert;
arXiv:0906.4127.

\bibitem{topologicalClifford}
H. Bombin, M.A. Martin-Delgado;
``Topological Quantum Distillation'';
Phys. Rev. Lett. {\bf 97}, 180501 (2006).
quant-ph/0605138.

\bibitem{topo3D}
H. Bombin, M.A. Martin-Delgado
``Exact Topological Quantum Order in D=3 and Beyond:
Branyons and Brane-Net Condensates'';
Phys. Rev. B {\bf 75}, 075103 (2007);
cond-mat/0607736.

\bibitem{tetraUQC}
H. Bombin, M.A. Martin-Delgado;
``Topological Computation without Braiding'';
Phys. Rev. Lett. {\bf 98}, 160502 (2007);
quant-ph/0610024.

\bibitem{wen_book}
X.G. Wen,  {\it Quantum Field Theory of Many-body Systems: From the Origin of Sound to an Origin of Light and Electrons} (Oxford Univ. Press, New York, 2004).

\bibitem{wen_fqh}
X.G. Wen,  
``Topological orders and edge excitations in fractional quantum Hall states'';
Adv. Phys., \textbf{44}, 405 (1995).

\bibitem{zoller05}
 A. Micheli, G.K. Brennen, P. Zoller,
``A toolbox for lattice spin models with polar molecules'';
Nat. Phys. {\bf 2}, 341 (2006).
quant-ph/0512222.


\bibitem{jiang08}
L. Jiang, G.K. Brennen, A.V. Gorshkov, K. Hammerer, M. Hafezi,
E. Demler, M.D. Lukin, P. Zoller,
``Anyonic interferometry and protected memories in atomic spin lattices'';
Nat. Phys. {\bf 4}, 482 (2008).
arXiv:0711.1365.


\bibitem{ol_review}
M. Lewenstein, A. Sanpera, V. Ahufinger, B. Damski, A. Sen, U. Sen, 
``Ultracold atomic gases in optical lattices: mimicking condensed matter 
physics and beyond'';
Adv. Phys. \textbf{ 56}, 243 (2007).




\bibitem{gottesman96}
D. Gottesman,
``Class of quantum error-correcting codes saturating the quantum Hamming bound''.
Phys. Rev. A {\bf 54}, 1862 (1996).


\bibitem{kitaev97}
A.\,Yu.\,Kitaev,
``Fault-tolerant quantum computation by anyons'',
Annals of Physics \textbf{303}, 2--30 (2003), quant-ph/9707021.


\bibitem{honeycomb}
A. Yu. Kitaev;
``Anyons in an exactly solved model and beyond'';
Ann. of Phys. {\bf 321}, 2--111 (2006).
arXiv:cond-mat/0506438v3.

\bibitem{duan03}
L.-M. Duan, E. Demler, M.D. Lukin,
``Controlling Spin Exchange Interactions of Ultracold Atoms 
in Optical Lattices'';
Phys. Rev. Lett.  {\bf 91},  090402 (2003).
arXiv:cond-mat/0210564.

\bibitem{levinwen05}
M. Levin, X.-G. Wen,
``String-net condensation:   A physical mechanism for topological phases''.
Phys. Rev. {\bf B 71}, 045110 (2005).

\bibitem{PCUTS}
C. Knetter, G. Uhrig; 
``Perturbation theory by flow equations: dimerized and frustrated S = 1/2 chain''; Eur. Phys. J. B{\bf 13}, 209 (2000).

\bibitem{wegner94}
F. Wegner,
``Flow-equations for Hamiltonians'';
Ann. der Phys.  {\bf 506}, 77 (1994).

\bibitem{honeycomb_bosonization}
J. Vidal, K.P. Schmidt, S. Dusuel;  
``Perturbative approach to an exactly solved problem: Kitaev honeycomb model'';
Phys. Rev. B {\bf 78},  245121 (2008).
arXiv:0809.1553v1.

\bibitem{laughlin83}
R.B. Laughlin,
``Anomalous quantum Hall effect - 
An incompressible quantum fluid with fractionally charged excitations'';
Phys. Rev. Lett.  {\bf 50},   1395 (1983).





\end{thebibliography}
\end{document}